\documentclass[aps,prd,fleqn,superscriptaddress]{revtex4}
\usepackage{graphicx,xcolor,natbib,braket,float}
\usepackage{amsmath,amssymb,amsfonts,cases}
\newcommand{\bse}{\begin{subequations}}
\newcommand{\ese}{\end{subequations}}
\newcommand{\be}{\begin{equation}}
\newcommand{\ee}{\end{equation}}
\newcommand{\bea}{\begin{eqnarray}}
\newcommand{\eea}{\end{eqnarray}}
\newcommand{\ba}{\begin{array}}
\newcommand{\ea}{\end{array}}

\usepackage[colorlinks=true, linkcolor=blue, bookmarks=true]{hyperref}

\begin{document}
\title{Probe Dependence of the Imaginary Part of HTEE}
\author{Mohammad Ali-Akbari\footnote{$\rm{m}_{-}$aliakbari@sbu.ac.ir}}
\affiliation{Department of Physics, Shahid Beheshti University, 1983969411, Tehran, Iran}

\begin{abstract}
We investigate the holographic timelike entanglement entropy (HTEE) in a five-dimensional anisotropic background, dual to a strongly coupled anisotropic plasma. Using the complex extremal surface method, we compute the HTEE analytically in the high-temperature, small-anisotropy limit $a/T \ll 1$. We consider two different orientations of the boundary timelike interval: one perpendicular to the anisotropy direction and one parallel to it. We find that the imaginary part of the HTEE is not a universal property of the geometry but depends sensitively on the orientation of the extremal surface relative to the anisotropy. This demonstrates that the imaginary part arising from the UV logarithmic divergence is a probe-dependent quantity. Our results suggest that the imaginary part of HTEE can serve as a diagnostic of the coupling between the extremal surface and the anisotropic degrees of freedom of the dual field theory.
\end{abstract}

\maketitle

\tableofcontents

\section{Introduction}
The AdS/CFT correspondence, in its strongest form, states that a supersymmetric Yang–Mills theory in \(3+1\) dimensions is dual to type IIB superstring theory on \(\mathrm{AdS}_5 \times S^5\), where the field theory lives on the conformal boundary of the bulk spacetime \cite{Maldacena:1997re, Witten:1998qj, Gubser:1998bc, Aharony:1999ti, DHoker:2002nbb, Casalderrey-Solana:2011dxg}. However, this duality is most commonly applied in a specific limit in which the field theory becomes strongly coupled and the string theory reduces to classical gravity. In this regime, the gauge theory is replaced by its strongly coupled version, while the string theory side is well approximated by classical supergravity. This provides a powerful tool for studying strongly coupled systems that are inaccessible via conventional perturbative methods in quantum field theory.
Moreover, the duality often provides a simple geometric counterpart for physical quantities and processes in the bulk, allowing them to be interpreted more clearly in terms of the dual field theory. For instance, the thermal properties of the plasma, such as temperature and entropy density, are encoded in the black hole geometry, while non-local observables like entanglement entropy are mapped to the areas of extremal surfaces in the bulk.

In quantum information theory, entanglement entropy quantifies the amount of quantum correlation between two subsystems \cite{Headrick:2019eth, Calabrese:2004eu, Casini:2009sr}. For a pure state $|\psi\rangle$ defined on a Hilbert space partitioned into a subsystem $A$ and its complement $A^c$, the entanglement entropy measures how much information is shared between the two parts. Equivalently, it quantifies the loss of information when one subsystem is traced out. It is defined as the von Neumann entropy of the reduced density matrix
\be
S_A = - \operatorname{Tr} \rho_A \ln \rho_A,
\ee
where $\rho_A = \operatorname{Tr}_{A^c} (|\psi\rangle \langle \psi|)$ is the reduced density matrix obtained by tracing over the complementary degrees of freedom.
Remarkably, this inherently quantum quantity has a simple and elegant geometric dual in the context of the AdS/CFT correspondence. According to the Ryu–Takayanagi (RT) proposal \cite{Ryu:2006bv, Hubeny:2007xt, Rangamani:2016dms, Nishioka:2009un}, the entanglement entropy of a boundary subsystem $A$ is given by the area of an extremal surface $\Gamma_A^s$ that extends into the bulk and anchors on the boundary such that its boundary coincides with that of $A$
\be\label{HEE}
S_A^{\rm HEE} = \frac{\operatorname{Area}(\Gamma_A^s)}{4G_N},
\ee
where $G_N$ is the five-dimensional Newton constant and $\Gamma^s$ denotes real spacelike extremal surface. This proposal has been extensively tested and applied to a wide variety of holographic systems, providing deep insights into the relationship between geometry and quantum entanglement.
It is important to emphasize that in the standard RT prescription, the extremal surface is real and spacelike and the resulting entanglement entropy is a real quantity.

Recently, following the success of holographic entanglement entropy, the HTEE has attracted significant attention \cite{Doi:2023zaf, Prihadi:2026nua, Basak:2023otu, Jena:2024tly}. Instead of considering a spacelike subsystem $A$ on a fixed time slice, HTEE explores quantum correlations between regions that are separated in time. This naturally raises the question of how quantum information is shared and evolves during the time evolution of the system. In the context of the AdS/CFT correspondence, there are three main approaches to compute the HTEE which we briefly summarize as follows:
\begin{itemize}
\item The first method, originally introduced in \cite{Doi:2022iyj, Doi:2023zaf}, consists of computing the standard holographic entanglement entropy for a spacelike subsystem and then analytically continuing the final result by replacing the spacelike interval length with $i$ times the timelike interval length. This approach is particularly simple and effective in two-dimensional conformal field theories where the entanglement entropy has a known universal form. In such cases, the analytic continuation yields a HTEE with a real part given by the usual logarithmic term and a universal imaginary part $i\pi/2$. However, this method has limitations. It works reliably only when the background metric has no off-diagonal components. In more general settings, with non-trivial off-diagonal metric components or in the presence of gravitational anomalies, the simple replacement $l \to i \Delta t$ does not yield the correct HTEE \cite{Doi:2023zaf, Ali-Akbari:2026xzx, Chu:2025sjv}. In particular, for boosted backgrounds, the analytic continuation from the spacelike result fails to capture the full structure of the HTEE.

\item The second approach, which is used in \cite{Afrasiar:2024ldn, Goki:2026hpl} and related works, is based on the construction of complex extremal surfaces in the bulk geometry. In this method, one first solves the extremal equation for a real extremal surface and expresses the turning point in terms of a conserved quantity, say \(C\). For a timelike boundary interval, the turning point becomes purely imaginary which corresponds to choosing \(C\) to be a purely complex number. This leads to a complex extremal surface that consists of two distinct parts:
\begin{itemize}
    \item A real part, which is obtained by taking the absolute value of the complex turning point and corresponds to a spacelike segment of the surface.
    \item A purely imaginary part, which arises from the imaginary turning point and corresponds to a timelike segment of the surface.
\end{itemize}
The full extremal surface is then formed by combining these two parts and the total area is obtained by summing their contributions. The real part of the area gives the usual logarithmic divergence and the finite real corrections, while the imaginary part yields a phase that is directly related to the timelike nature of the boundary interval.

\item The third approach, which from our point of view is the most natural and conceptually cleanest one, starts directly with a complexified bulk geometry. In this method, one does not begin with a real extremal surface and then analytically continue it. Instead, the bulk spacetime itself is treated as complex from the outset and the extremal surface is computed entirely within this complexified geometry.
The extremal surface \(\Gamma_A^t\) is then substituted directly into the holographic formula \cite{Heller:2024whi}
\be
S_A^{{\rm HTEE}} = \frac{\operatorname{Area}(\Gamma_A^t)}{4G_N},
\ee
where \(\Gamma_A^t\) denotes the complex extremal surface anchored to the timelike boundary interval. This complex surface, when evaluated, naturally yields both a real and an imaginary part for the HTEE without the need for any manipulation of the conserved momentum, analytic continuation of the interval or post-processing of the result.
The key advantage of this method is that it treats the complexification of the geometry as a fundamental ingredient from the very beginning which is conceptually consistent with the timelike nature of the boundary interval. It provides a direct geometric interpretation of both the real and imaginary components of the HTEE. 
\end{itemize}

In this work, we adopt this third approach. We work directly in the complexified five-dimensional anisotropic background, compute the complex extremal surface and obtain the HTEE. This allows us to identify the origin of the imaginary part of the HTEE as arising from the complex nature of the turning point and the integration contour in the complexified bulk geometry.

\section{Anisotropic background metric}
To compute the HTEE, we consider the following five-dimensional anisotropic metric \cite{Mateos:2011tv}
\be\label{metric}
ds^{2} = \hat{g}_{tt} \, dt^{2} + \hat{g}_{uu} \, du^{2} + \hat{g}_{zz} \, dz^{2} + \hat{g}_{ss} \, (dx^{2} + dy^{2}),
\ee
with the metric components given by
\bse\label{component}
\begin{align}
\hat{g}_{tt} &= -e^{-\phi/2}u^{-2} \mathcal{F} \mathcal{B}=e^{-\phi/2}g_{tt}, \label{eq:gtt} \\
\hat{g}_{uu} &= e^{-\phi/2}u^{-2} \mathcal{F}^{-1}=e^{-\phi/2}g_{uu}, \label{eq:guu} \\
\hat{g}_{zz} &= e^{-\phi/2}u^{-2} \mathcal{H}=e^{-\phi/2}g_{zz}, \label{eq:gzz} \\
\hat{g}_{ss} &= e^{-\phi/2}u^{-2}=e^{-\phi/2}g_{ss}, \label{eq:gss}
\end{align}
\ese
where $\mathcal{H} = e^{-\phi}$.
The functions $\mathcal{F}$, $\mathcal{B}$, and $\phi$ are expanded in the high-temperature, small-anisotropy limit $a/T \ll 1$ as
\bse\label{metriccom}
\begin{align}
\mathcal{F} &= 1 - \frac{u^{4}}{u_{h}^{4}} + a^{2} \mathcal{F}_{2} 
= 1 - \frac{u^{4}}{u_{h}^{4}} + \frac{a^{2}}{24 u_{h}^{2}} \left[ 8u^{2}(u_{h}^{2} - u^{2}) - 10 u^{4} \log 2 + (3u_{h}^{4} + 7u^{4}) \log\!\left(1 + \frac{u^{2}}{u_{h}^{2}}\right) \right], \label{eq:F2} \\
\mathcal{B} &= 1 + a^{2} \mathcal{B}_{2} 
= 1 - \frac{a^{2} u_{h}^{2}}{24} \left[ \frac{10u^{2}}{u^{2} + u_{h}^{2}} + \log\!\left(1 + \frac{u^{2}}{u_{h}^{2}}\right) \right], \label{eq:B2} \\
\phi &= a^{2} \phi_{2} = -\frac{a^{2} u_{h}^{2}}{4} \log\!\left(1 + \frac{u^{2}}{u_{h}^{2}}\right), \label{eq:phi2}
\end{align}
\ese
where the parameter $a$ is the anisotropy parameter.
These expansions are exact to order $a^2$ in the metric functions. The explicit $a^4$ terms in $\mathcal{F}, \mathcal{B}, \phi$ are not shown here because they do not contribute to the leading $a^2$ corrections to the HTEE that we compute in this paper. All calculations in the following sections are performed consistently to order $a^2$ in the final entropy and we will not explicitly state the perturbative order at every step.

This background was originally introduced in \cite{Mateos:2011tv} as a solution of type IIB string theory and is holographically dual to a strongly coupled, anisotropic Yang-Mills plasma at finite temperature. The metric preserves translational invariance along the boundary directions but breaks the rotational symmetry between the $z$ and the $x,y$ directions whenever $\mathcal{H} \neq 1$. The breaking of $SO(3)$ rotational symmetry is controlled by the anisotropy parameter $a$ which sources a pressure anisotropy in the dual field theory. In the limit $a \to 0$, the metric reduces to the standard Schwarzschild-AdS$_5$ black hole corresponding to an isotropic plasma.
The radial coordinate $u$ runs from the boundary at $u=0$ to the horizon at $u=u_{h}$, where $g_{tt}(u_{h}) = 0 = g_{uu}^{-1}(u_{h})$. We work in the high-temperature regime $a/T \ll 1$, for which the temperature is given by
\be
T = \frac{1}{\pi u_{h}} + \frac{u_{h}(5\log 2 - 2)}{48\pi} \, a^{2}.
\ee

For arbitrary values of the anisotropy parameter and temperature, the background is known only numerically. However, since our goal is to compute the HTEE analytically, we restrict ourselves to the high-temperature, small-anisotropy limit where the above expansions are valid. In this regime, the anisotropy enters as a small perturbation, allowing us to compute the HTEE perturbatively in $a^2$ while keeping the time interval $\Delta t$ arbitrary within the small-$(T\Delta t)$ limit. A detailed investigation of the properties of this anisotropic background has been carried out in the literature, covering its thermodynamic stability, phase structure, transport coefficients and the behavior of non-local observables such as entanglement entropy and Wilson loops; see for example \cite{Mateos:2011tv, Rahimi:2018ica, Ali-Akbari:2014nua, Ali-Akbari:2014xea, Ali-Akbari:2013txa, Rebhan:2011vd, Ge:2014aza}.

Finally, it is important to note that the metric \eqref{metric} and its components \eqref{component} are written in the Einstein frame, where the factor \(e^{-\phi/2}\) in each component reflects the difference between the Einstein (E) and string (S) frames. Consequently, an extra factor \(e^{-3\phi/4}\) appears in the area functional, as shown in \eqref{area1} and \eqref{area2}. It is easy to see
\be 
\sqrt{-\det(\hat{g}^E_{\mu\nu})}=e^{-3\phi/4}\sqrt{-\det(g^S_{\mu\nu})}.
\ee

\section{Holographic timelike entanglement entropy}
In order to compute the HTEE, we consider two different orientations for the timelike strip-like subsystem $A$, defined by
\be\label{interval}
\Delta t = 
\begin{cases}
(1)\ \ A_x=\{(t,0,y,z): |t|\le\frac{\Delta t}{2},\ x = 0,\ (y,z)\in \mathbb{R}^2 \},\ \ {\rm Parallel}, \\[6pt]
(2)\ \ A_z= \{(t,x,y,0) :|t|\le\frac{\Delta t}{2},\ (x,y)\in \mathbb{R}^2,\ z = 0\},\ \ {\rm Perpendicular},
\end{cases}
\ee
where in both cases the interval extends along the transverse directions (i.e., it is a strip of infinite extent in the spatial directions orthogonal to the chosen coordinate).
The first choice places the interval parallel to the anisotropy direction (the $z$-direction), while the second places it perpendicular to it. This distinction is physically meaningful because the background metric \eqref{metric} breaks the rotational symmetry between the $z$ and the $x,y$ directions, as seen from the fact that $g_{zz} \neq g_{ss}$ when $\mathcal{H} \neq 1$. There remains a $U(1)$ symmetry between the $x$ and $y$ directions, so choosing the interval along $x$ or $y$ is equivalent; however, the $x$ and $z$ directions are not equivalent due to the anisotropy parameter $a$.

By studying these two configurations, we can probe the anisotropic response of the dual field theory. The extremal surfaces anchored to these intervals will explore different combinations of the metric components, leading to different $a^2$ corrections in the HTEE. This allows us to isolate the effect of the anisotropy and to determine whether the imaginary part of the HTEE is a universal property of the geometry or rather a probe-dependent quantity.

In order to perform the HTEE calculation analytically, we need a second approximation, namely the small-time interval limit $T \Delta t \ll 1$, which corresponds to $|u_*| \ll u_h$. Here, $u_*$ is the turning point of the extremal surface in the complexified bulk geometry. The condition $|u_*| \ll u_h$ implies that the extremal surface remains close to the boundary and does not penetrate deeply into the bulk, thus probing only the UV (near-boundary) region of the geometry. As a consequence, the effect of the black hole horizon (and hence the temperature) is eliminated from the leading-order results. Temperature effects would appear only at subleading order which we neglect in our perturbative expansion.
In this limit, the metric components \eqref{metriccom} reduce to the simplified form
\bse\label{ametric}
\begin{align}
\mathcal{F} &= 1 - \frac{u^4}{u_h^4} + \frac{11}{24} a^2 u^2 + \frac{\gamma_F}{24 u_h^2} a^2 u^4, \label{eq:F_small} \\
\mathcal{B} &= 1 - \frac{11}{24} a^2 u^2 + \frac{7}{16 u_h^2} a^2 u^4, \label{eq:B_small} \\
\mathcal{H} &= 1 + \frac{1}{4} a^2 u^2 - \frac{1}{8 u_h^2} a^2 u^4, \label{eq:H_small}
\end{align}
\ese
where the constant $-\gamma_F= 10 \log 2 + 8 + \frac{3}{2}$.
The expansion above keeps terms up to order $a^2 u^4$ in the metric functions which is sufficient for computing the leading $a^2$ corrections to the HTEE. Note that the $u^4/u_h^4$ term in $\mathcal{F}$ is kept separately, as it contributes to the subleading temperature-dependent corrections. However, as argued above, these terms will not affect the leading $1/(\Delta t)^2$ and constant $a^2$ contributions in the small-$(T \Delta t)$ limit. This is consistent with the fact that the leading UV behaviour of the HTEE is insensitive to the IR structure of the geometry. In short, in this paper we consider the limit
\be
T\Delta t \ll \frac{a}{T} \ll 1 .
\ee

\subsection{HTEE for the interval at $x=0$}
Now let us compute the area functional, corresponding to configuration $(1)$ in \eqref{interval}, that must be extremalized in order to obtain the HTEE. For this timelike interval, the area of the extremal surface is given by
\be\label{area1}
\frac{\Gamma(A_x)}{2}\equiv \frac{{\rm Area\left(\Gamma^t_{A_x}\right)}}{2} = V_{2x} \int_{u_*}^{\epsilon} du \, e^{-3\phi/4}\sqrt{ g_{zz} g_{ss} \left( g_{tt} \dot{t}^2 + g_{uu} \right) }=V_{2x}\int_{u_*}^{\epsilon} {\cal L}_x\ du,
\ee
where $V_{2x} = \int dy \, dz$ is the volume of the transverse directions and $\dot{t} \equiv dt/du$ denotes the derivative with respect to the holographic radial coordinate. The UV cutoff is denoted by $\epsilon$, which is a real number, although the radial coordinate $u$ is generally complex because we are working in a complexified bulk geometry.

Since the metric components depend only on $u$, the coordinate $t$ is cyclic. The corresponding conserved momentum $E_x$ is therefore given by
\be
E_x = \frac{\partial \mathcal{L}_x}{\partial \dot{t}}.
\ee
Solving for $\dot{t}$ yields
\be \label{dot}
\dot{t}^2(u) = \frac{E_x^2 g_{uu}}{g_{tt} \left(g_{tt} g_{zz} g_{ss}e^{-3\phi/2} - E_x^2 \right)}.
\ee
The turning point $u_*$ of the extremal surface is defined by the condition $\dot{t} \to \infty$, which corresponds to the vanishing of the denominator in \eqref{dot}:
\be\label{turning}
\mathcal{K}_* u_*^{-6} + E_x^2 = 0,
\ee
where we have introduced $\mathcal{K} \equiv \mathcal{F} \mathcal{B} \mathcal{H}^{\frac{5}{2}}$ and $\mathcal{K}_* \equiv \mathcal{K}(u_*)$. At this point, the extremal surface reaches its deepest radial penetration before returning to the boundary.
Substituting the turning point condition \eqref{turning} into \eqref{dot}, we obtain 
\be\label{dot2}
\dot{t}^2(u) = \frac{u_*^{-6} \mathcal{K}_*}{\mathcal{F}^2 \mathcal{B} \left( u_*^{-6} \mathcal{K}_* - u^{-6} \mathcal{K} \right)}.
\ee

To capture the timelike nature of the boundary interval, we work directly in the complexified bulk geometry and choose a purely imaginary turning point by setting
\be\label{change0}
u = i s, \qquad u_* = i s_*, 
\ee
and define the dimensionless real parameter $r \equiv \frac{u}{u_*}$.
At this stage, two important points are worth mentioning.
\begin{itemize}
\item
On the field theory side, real parameters such as the temperature and the UV cutoff correspond to different points along the radial direction, since the latter is dual to the energy scale of the dual field theory. Consequently, these points on the radial direction must be real, leading to real values of the temperature and UV cutoff. However, under the change of variables \eqref{change0}, these special points become complex. For illustration, the contour is plotted in figure \ref{fig}. In this paper, since the temperature dependence drops out in the limits we consider, we focus solely on the UV cutoff. The spacetime boundary is located at \(u = \epsilon\), and along the complex path \(u = i s_* r\), this boundary corresponds to
\be\label{change}
r = \frac{\epsilon}{i s_*} = -i \frac{\epsilon}{s_*}.
\ee
We denote this complex lower limit by \(\epsilon^* \equiv -i \epsilon / s_*\), so that the integration over \(r\) runs from \(\epsilon^*\) to \(1\). It is important to note that when the integrals over \(r\) contain a logarithmic divergence at \(r=0\), the complex lower limit \(\epsilon^*\) cannot be simply set to zero. The complex logarithm evaluated at this point gives
\be\label{complog}
\ln \epsilon^* = \ln\left(-i \frac{\epsilon}{s_*}\right) = \ln\left(\frac{\epsilon}{s_*}\right) - \frac{i\pi}{2}.
\ee
This phase is the origin of the imaginary part of the HTEE. In contrast, if the integral is convergent at \(r=0\), the limit \(\epsilon^* \to 0\) can be taken safely and no imaginary contribution arises.
Therefore, the imaginary part of the HTEE is a direct consequence of the complexification of the bulk geometry required for timelike boundary intervals, combined with the presence of UV logarithmic divergences in the area functional, as we will see later. The complex nature of the turning point and the integration path is a necessary ingredient and the resulting imaginary part is a genuine feature of the timelike setup.

\begin{figure}[H]
  \centering
  \includegraphics[width=0.55\textwidth]{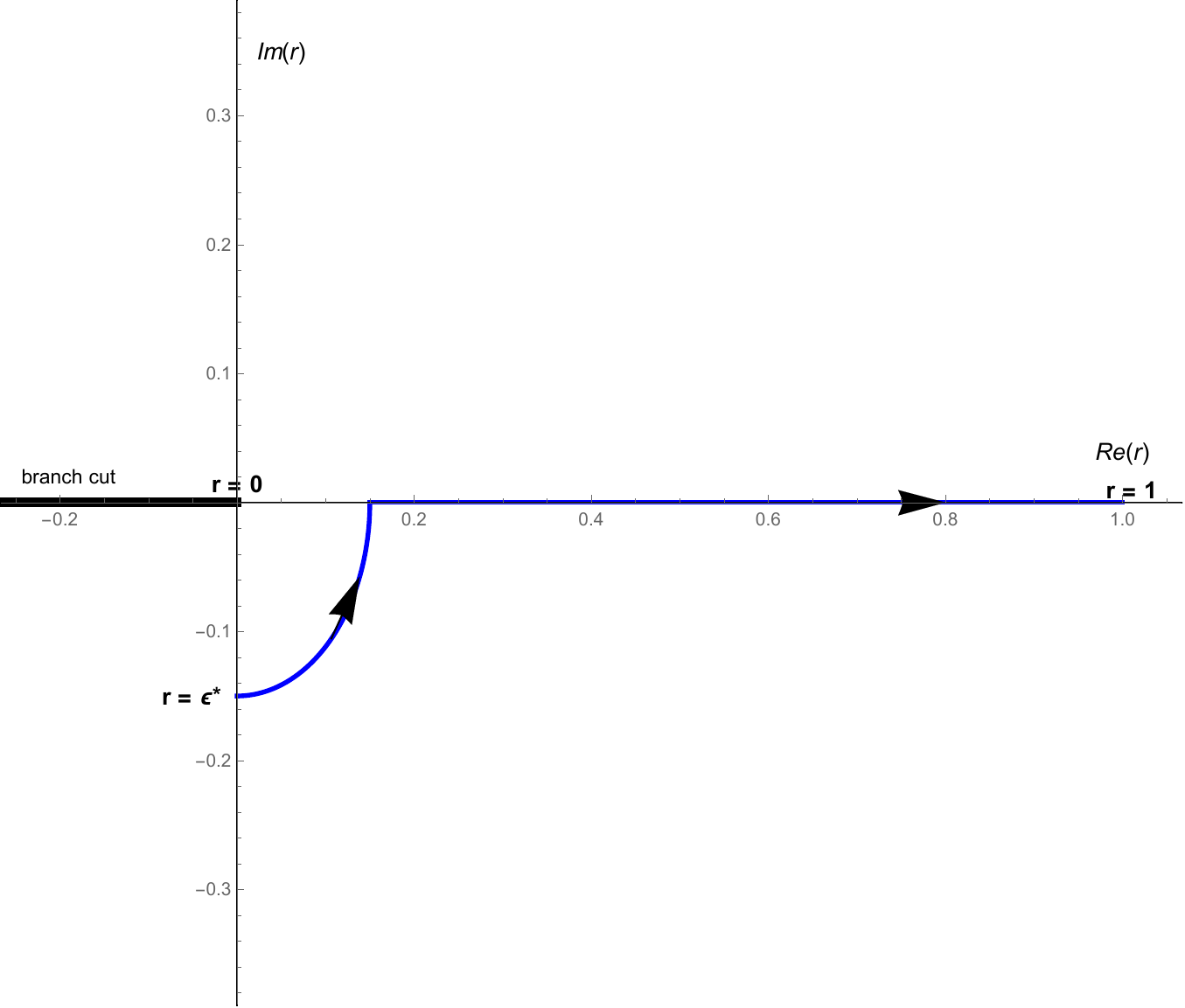}
 \caption{The integration contour in the complex $r$-plane. It consists of two parts: it starts at the purely imaginary point $\epsilon^*$ on the imaginary axis, follows a small arc of radius $|\epsilon^*|$ to $r=|\epsilon^*|$ on the real axis, avoiding the branch point at $r=0$, and then continues along the real interval to $r=1$. The branch cut is taken along the negative real axis, corresponding to the principal branch of the logarithm, so that $\ln\epsilon^* = \ln(\epsilon/s_*) - i\pi/2$, consistent with \eqref{complog}.}
  \label{fig}
\end{figure}

\item 
In fact, in \eqref{change0}, one may alternatively choose \(u = -i s\) and \(u_* = -i s_*\). This choice leaves the time interval \(\Delta t\) and the real parts of the areas \(\Gamma(A_x)\) and \(\Gamma(A_z)\) unchanged (as we will demonstrate in the next subsection). The only difference appears in the imaginary parts of \(\Gamma(A_x)\) and \(\Gamma(A_z)\), whose signs flip from negative (positive) to positive (negative). Accordingly, in the remainder of this work, we adopt the plus-sign convention, though this ambiguity remains unresolved.
\end{itemize}

With this parametrization, \eqref{dot2} becomes
\be\label{dot3}
\dot{t}(i s_* r) = \frac{\sqrt{\mathcal{K}_*} \, r^3}{i \, \mathcal{F}(i s_* r) \sqrt{\mathcal{B}(i s_* r)} \, \sqrt{ \mathcal{K}(i s_* r) - \mathcal{K}_* r^6 }},
\ee
where now $\mathcal{K}_* \equiv \mathcal{K}(i s_*)$. The factor of $i$ in the denominator reflects the complexified bulk geometry which is necessary for timelike boundary intervals.

The boundary time interval $\Delta t$ is obtained by integrating along the contour $u = i s_* r$
\be
\frac{\Delta t}{2} = i s_* \int_{\epsilon^*}^1 dr \, \dot{t}(i s_* r).
\ee
This integral yields a real, positive $\Delta t$, as we will demonstrate in the following.
To proceed analytically, we expand the metric functions $\mathcal{F}$, $\mathcal{B}$ and $\mathcal{H}$ on the complex contour $u = i s_* r$ to order $a^2$. Using the small-$|u|$ expansions in \eqref{ametric}, we obtain
\be
\frac{\sqrt{\mathcal{K}_*}}{\mathcal{F}(i s_* r) \sqrt{\mathcal{B}(i s_* r)}} = 1 + a^2 s_*^2 \left( \frac{11}{48} r^2 - \frac{5}{16} \right) + s_*^4 \left[ \frac{1}{u_h^4}\left(r^4-\frac{1}{2}\right)+ \left( \frac{1}{16}+\frac{\gamma_F}{48} -[\frac{7}{32}+\frac{\gamma_F}{24}]r^4 \right) \frac{a^2}{u_h^2} \right],
\ee
and
\be
\begin{split}
\frac{1}{\sqrt{\mathcal{K}(i s_* r) - \mathcal{K}_* r^6}}
= \frac{1}{\sqrt{1 - r^6}} \left( 1 + \frac{5}{16} a^2 s_*^2 \frac{r^2 - r^6}{1 - r^6} \right).
\end{split}
\ee
Multiplying the two expansions and integrating over $r$, we find that the time interval takes the compact form
\be
\frac{\Delta t}{2} = s_* \left[ \int_{0}^{1} \frac{r^3 \, dr}{\sqrt{1 - r^6}} + a^2 s_*^2 \left( \frac{11}{48} \int_0^1 \frac{r^5 \, dr}{\sqrt{1 - r^6}} - \frac{5}{16} \int_0^1 \frac{r^3 \, dr}{\sqrt{1 - r^6}} + \frac{5}{16} \int_0^1 \frac{r^3(r^2 - r^6) \, dr}{(1 - r^6)^{3/2}} \right) \right].
\ee
The integrals are all elementary and can be expressed in terms of Beta functions. 
Evaluating the integrals, we obtain the relation between the boundary time interval $\Delta t$ and the real turning point parameter $s_*$ as
\be\label{deltats}
\frac{\Delta t}{2} = C_1 s_* + C_2 a^2 s_*^3,
\ee
where the constants $C_1$ and $C_2$ are given by
\be\begin{split} 
C_1&=\frac{1}{6}B\left(\frac{2}{3},\frac{1}{2}\right),\\
C_2&=\frac{1}{6}\left[\frac{11}{48}B\left(1,\frac{1}{2}\right)-\frac{5}{16}B\left(\frac{2}{3},\frac{1}{2}\right)-\frac{5}{16}B\left(\frac{5}{3},-\frac{1}{2}\right)+\frac{5}{16}B\left(1,-\frac{1}{2}\right)\right].
\end{split}\ee
This relation will be inverted later to express $s_*$ in terms of the physical time interval $\Delta t$.

Similarly, one can obtain the on-shell area by substituting the complexified solution \eqref{dot2} into the area functional. The half-area (due to the symmetry of the extremal surface) is given by
\be
\frac{\Gamma(A_x)}{2V_{2x}} = \int_{u_*}^{\epsilon} du \, \frac{\mathcal{H}^{\frac{5}{2}}}{u^3} \sqrt{ \frac{\mathcal{B}}{\mathcal{K}(u) - \mathcal{K}_* (u/u_*)^6 } },
\ee
where the integration runs from the turning point $u_*$ to the boundary.
After changing variables to $u = i s_* r$ and expanding the metric functions to order $a^2$, the half-area takes the form
\be\label{intA}
-\frac{\Gamma(A_x)}{2V_{2x}} = \frac{1}{s_*^2} \int_{\epsilon^*}^1 \frac{dr}{r^3 \sqrt{1 - r^6}} + a^2 \left[ \frac{5}{16} \underbrace{\int_{\epsilon^*}^1 \frac{(1-r^4)dr}{r (1 - r^6)^{3/2}}}_{I_2} - \frac{19}{48} \underbrace{\int_{\epsilon^*}^1\frac{dr}{r \sqrt{1 - r^6}}}_{I_3} \right],
\ee
where the overall minus sign on the left-hand side comes from the orientation of the integral. 
Performing the integrals, we find
\be\label{areax}
\frac{\Gamma(A_x)}{2V_{2x}} = -\frac{C_3}{s_*^2} - C_4 a^2,
\ee
where the constants $C_3$ and $C_4$ are given by
\bse
\begin{align}
C_3 &= \frac{1}{6} B\!\left( -\frac{1}{3}, \frac{1}{2} \right) = -\frac{1}{12} B\!\left( \frac{2}{3}, \frac{1}{2} \right), \\
\label{29b} C_4 &= C_4^x-\frac{i\pi}{24}-\frac{1}{12}\ln(s_*),\\
C_4^x&=-\frac{1}{6} \left[ \frac{\ln 2}{6}+\frac{5}{8} +\frac{5}{16} B\!\left( \frac{2}{3}, -\frac{1}{2} \right) \right].
\end{align}
\ese
Note that the integrals \(I_2\) and \(I_3\) in \eqref{intA} are divergent as \(\epsilon \to 0\) and therefore require regularization; the details are provided in Appendix \ref{appa}. Upon applying the regularization procedure with the complex cutoff \(\epsilon^* = -i\epsilon/s_*\) introduced in \eqref{complog}, the imaginary part of the on-shell area is found to be proportional to \(\frac{i\pi}{2}\). This imaginary contribution originates from the logarithmic UV divergence, whose complex lower limit generates a nontrivial phase.

Having obtained the relation between the boundary time interval \(\Delta t\) and the real turning point parameter \(s_*\) in \eqref{deltats}, we now invert it to express \(s_*\) in terms of \(\Delta t\). Solving perturbatively to first order in \(a^2\), we find
\be
s_* = \frac{\Delta t}{2 C_1} - \frac{C_2}{8 C_1^4} a^2 (\Delta t)^3.
\ee
This inversion is valid in the small-\(a^2\) limit and captures the leading anisotropic correction to the turning point.
Substituting this expression into the on-shell area in \eqref{areax} and using \eqref{29b}, we obtain the final result for the HTEE area as
\be
\frac{\Gamma(A_z)}{V_{2x}} = \frac{4 C_1^3}{(\Delta t)^2} + 2 \left( C_2 - C^x_4 +\frac{1}{12}\ln\left(\frac{\Delta t}{2C_1}\right) + \frac{i\pi}{24} \right) a^2.
\ee
The real part of the area
\be
{\rm Re}\left(\frac{\Gamma(A_z)}{V_{2x}}\right) = \frac{4 C_1^3}{(\Delta t)^2} + 2 \left[ C_2 - C^x_4 +\frac{1}{12}\ln\left(\frac{\Delta t}{2C_1}\right)\right] a^2,
\ee
contains the leading isotropic contribution \(\frac{4 C_1^3}{(\Delta t)^2}\), which matches the known result for a timelike interval in AdS\(_5\), for instance see \cite{Gong:2025pnu}. The coefficient \(C_1^3 \approx 0.080166\) gives \(4 C_1^3 \approx 0.320664\), which agrees with the numerical value \(0.321\) reported in the literature.
The imaginary part is given by
\be
 {\rm Im}\left(\frac{\Gamma(A_z)}{V_{2x}}\right) = \frac{\pi}{12} a^2.
\ee
It is important to emphasize that the coefficient \(\frac{1}{12}\) in the imaginary part is directly linked to the coefficient of the logarithmic UV divergence which depends on how the extremal surface couples to the anisotropic metric component \(\mathcal{H}\). As we will see in the next section, choosing a different orientation of the boundary interval (e.g., at \(z=0\)) yields a different coefficient, confirming that this imaginary part is not a universal geometric property but rather a probe-dependent quantity.

\subsection{HTEE for the interval at $z=0$}
For comparison, we now consider the alternative orientation where the timelike interval is placed at \(z=0\) instead of \(x=0\), the second case in \eqref{interval}. In this case, the extremal surface extends along the \(x\) and \(y\) directions, both of which have the same metric component \(g_{ss}\). The transverse volume factor is therefore \(g_{ss}\) rather than \(\sqrt{g_{zz}g_{ss}}\), and the area functional becomes
\be\label{area2}
\frac{\Gamma(A_z)}{2}\equiv \frac{{\rm Area\left(\Gamma^t_{A_z}\right)}}{2} = V_{2z} \int_{u_*}^\epsilon du \,e^{-3\phi/4} g_{ss} \sqrt{ g_{tt} \dot{t}^2 + g_{uu} }=V_{2z} \int_{u_*}^\epsilon {\cal L}_z\ du,
\ee
with $V_{2z}=\int dx dy$. The corresponding conserved momentum and turning point condition are
\begin{align}
E_z &=\frac{\partial {\cal L}_z}{\partial \dot{t}},\\
\dot{t}^2(u) &= \frac{E_z^2 g_{uu}}{g_{tt} (g_{ss}^2 g_{tt}e^{-3\phi/2} - E_z^2)},\\
(g_{ss}^2 g_{tt} & e^{-3\phi/2})_* - E_z^2=0.
\end{align}
Following the same steps as in the previous section, we obtain the time interval relation
\be
\frac{\Delta t}{2} = C_1 s_* + C_6 a^2 s_*^3,
\ee
where the constants are now
\be
C_6 = \frac{1}{6} \left[\frac{11}{48}B\left(1,\frac{1}{2}\right)-\frac{3}{16}B\left(\frac{2}{3},-\frac{1}{2}\right)+\frac{3}{16}B\left(1,-\frac{1}{2}\right)\right].
\ee

The half-area takes the simpler form
\be
\frac{\Gamma(A_z)}{2V_{2z}} = -\frac{C_3}{s_*^2} - C_8 a^2,
\ee
with
\be
\begin{split}
C_8 &= C_8^z+\frac{5i\pi}{24}+\frac{5}{12}\ln(s_*) ,\\
C_8^z&=\frac{1}{6} \left[ \frac{5\ln 2}{6} - \frac{3}{16} B\!\left( \frac{2}{3}, -\frac{1}{2} \right) \right].
\end{split}
\ee

Inverting the relation for \(s_*\) and substituting into the half-area yields the final on-shell area for the interval at \(z=0\)
\be
\frac{\Gamma(A_z)}{V_{2z}} = \frac{4 C_1^3}{(\Delta t)^2} + 2 \left( C_6 - C^z_8 - \frac{5}{12}\ln\left(\frac{\Delta t}{2C_1}\right)- \frac{5i\pi}{24} \right) a^2.
\ee

The imaginary part of this expression is
\be
{\rm Im}\left(\frac{\Gamma(A_z)}{V_{2z}}\right) = - \frac{5\pi}{12} a^2.
\ee

\section{Concluding Remarks}
Our final results are 
\be\begin{split}
\frac{4G_N}{V_{2x}}S_{\Delta t}^{{\rm HTEE}}& = 
\frac{4 C_1^3}{(\Delta t)^2} + 2 \left[C_2 - C_4^x+\frac{1}{12}\ln\left(\frac{\Delta t}{2C_1}\right)\right] a^2+\frac{i\pi}{12} a^2,\cr
\frac{4G_N}{V_{2z}}S_{\Delta t}^{{\rm HTEE}}&=\frac{4 C_1^3}{(\Delta t)^2} + 2\left[C_6 - C_8^z-\frac{5}{12}\ln\left(\frac{\Delta t}{2C_1}\right)\right] a^2-\frac{5i\pi}{12} a^2.
\end{split}\ee
Before closing this paper, we would like to emphasize the following key points regarding the HTEE in the anisotropic background:

\begin{itemize}
\item The leading terms of the real part of the HTEE are independent of the orientation of the timelike interval and are identical for both probes. In the small-$(T\Delta t)$ limit, they are given by $\frac{4C_1^3}{(\Delta t)^2}$, which matches the known AdS$_5$ result. However, the subleading real terms are orientation-dependent and receive non-trivial corrections proportional to the anisotropy parameter $a^2$.

\item For both orientations, the leading imaginary part of the HTEE vanishes. This is consistent with the fact that in AdS$_5$ (pure or thermal), the HTEE is purely real. The imaginary part only appears when the anisotropy is turned on, i.e., when $a \neq 0$, and it is proportional to $a^2$, in the limit of $a/T\ll 1$.

\item Unlike the well-studied two-dimensional strongly coupled field theories, where the imaginary part of the HTEE is constant, positive and independent of temperature and boost velocity, here the imaginary part is
\begin{itemize}
\item non-constant,
\item positive or negative
\item orientation-dependent,
\item proportional to the square of the anisotropy parameter, when $a/T\ll 1$.
\end{itemize}
This is a direct consequence of the fact that the imaginary part originates from the UV logarithmic divergence whose coefficient depends on the specific metric components that the extremal surface couples to.

\item For both orientations, the absolute value of the imaginary part increases monotonically with the anisotropy parameter $a$ in the limit of $a/T\ll 1$. This indicates that the anisotropy enhances the imaginary contribution to the HTEE, making it more significant as the pressure anisotropy of the dual plasma grows.

\item
The magnitude of the imaginary part is larger for the interval along the $z$-direction (parallel to the anisotropy) than for the corresponding interval along the $x$-direction (perpendicular to it). Explicitly, we obtain
\be
\left|\operatorname{Im}\left(\frac{\Gamma(A_x)}{V_{2x}}\right)\right| < \left|\operatorname{Im}\left(\frac{\Gamma(A_z)}{V_{2z}}\right)\right|.
\ee
This indicates that the imaginary part of the HTEE is a sensitive probe of the underlying anisotropic structure. We emphasize that this conclusion is independent of the sign ambiguity discussed before.

\item 
Finally, our results demonstrate that the imaginary part of the HTEE is not a universal geometric property of the spacetime. Instead, it is a probe-dependent quantity that reflects the specific coupling of the extremal surface to the background anisotropy. 
\end{itemize}


\section*{Acknowledgments}
We are sincerely grateful to the referee for his insightful comments which greatly improved the presentation of this manuscript. We also thank M.~M.~Daryaei Goki for useful discussions and feedback and DeepSeek for its assistance in refining the text.

\appendix
\section{Regularization of divergent integrals}\label{appa}
The second and third integrals in \eqref{intA} contain divergences. Note that in \(I_2\), the integral
\be
\int_{0}^{1}\frac{r^3\,dr}{(1-r^6)^{3/2}}=\frac{1}{6}B\left(\frac{2}{3},-\frac{1}{2}\right)
\ee
is finite in the limit \(\epsilon\rightarrow 0\). Therefore, to obtain the finite parts of the remaining integrals, we use
\begin{equation}
\begin{split}
I_2&=\int_{\epsilon^*}^{1-\delta}\frac{dr}{r(1-r^6)^{3/2}}
    =\frac{1}{6}\left[\frac{2}{\sqrt{1-r^6}}
    -\ln\left|\frac{1+\sqrt{1-r^6}}{1-\sqrt{1-r^6}}\right|\right]_{\epsilon^*}^{1-\delta},\\
&=\frac{1}{3\sqrt{6\delta}}-\ln\epsilon^*+\frac{\ln4}{6}-\frac{1}{3},\\
&=\frac{1}{3\sqrt{6\delta}}-\ln\epsilon+\frac{i\pi}{2}+\frac{\ln2}{3}-\frac{1}{3}+\ln(s_*).
\end{split}
\end{equation}
where $\delta\rightarrow 0$ and
\begin{equation}
\begin{split}
I_3&=\int_{\epsilon^*}^1\frac{dr}{r(1-r^6)^{1/2}}
    =\frac{1}{6}\ln\left|\frac{1-\sqrt{1-r^6}}{1+\sqrt{1-r^6}}\right|_{\epsilon^*}^1,\\
&=-\ln\epsilon_*+\frac{\ln4}{6},\\
&=  - \ln\epsilon +\frac{i\pi}{2}+\frac{\ln2}{3}+\ln(s_*).
\end{split}
\end{equation}
It is clear that \(I_2\) has both UV and IR divergences, whereas \(I_3\) has only a UV divergence. These divergences can be eliminated by adding appropriate counterterms, so we retain only the finite terms in the related HTEE.

\end{document}